\documentstyle[twocolumn,prl,aps,epsfig,amssymb]{revtex}
\def\beqn{\begin{eqnarray}}
\def\eeqn{\end{eqnarray}}
\relax
\newcommand{\ba}[1]{\begin{array}{#1}}
\def\ea{\end{array}}

\def\beq{\begin{equation}}
\def\eeq{\end{equation}}
\def\bea{\begin{eqnarray}}
\def\eea{\end{eqnarray}}

\def\dis{\displaystyle}
\def\f{\frac}
\def\[{\left[}
\def\]{\right]}
\def\({\left(}
\def\){\right)}

\def\GL{{\f{\lambda^a}{2}}}

\def\cut{{\Lambda}}
\def\HH{{\widetilde{H}}}
\def\K{\kappa}
\def\ov{\overline}

\def\U1em{{U(1)_{\rm em}}}
\def\ra{\rightarrow}
\def\T{{\cal T}}

\def\B{{\cal B}}

\def\G{{\cal G}}
\def\GSM{{\cal G}_{\rm SM}}

\def\ttbar{{t\bar{t}}}

\def\sq2{\sqrt{2}}

\def\STU{{(S,T,U)}}
\def\Zbb{{Zb\bar{b}}}
\def\ca{c_{\alpha}}
\def\sa{s_{\alpha}}
\def\End{\end{document}}
\begin{document}                                                              
\draft

\twocolumn[\hsize\textwidth\columnwidth\hsize\csname
@twocolumnfalse\endcsname
     
\title{   
       New Topflavor Models with Seesaw Mechanism 
}  
\author{{\sc Hong-Jian He}\,$^1$,~~ 
        {\sc Tim M.P. Tait}\,$^2$,~~ 
        {\sc C.--P. Yuan}\,$^{1,3}$
}
%\address{\phantom{ll}}
\address{
\vspace*{2mm}
$^1$Michigan State University, East Lansing, Michigan 48824, USA\\
$^2$Argonne National Laboratory, Argonne, Illinois 60439, USA\\
$^3$CERN, CH-1211, Geneva, Switzerland
}
%\date{\thisday}
\maketitle
\begin{abstract}
\hspace*{-0.35cm}
A new class of models are constructed in which the third family
quarks, but not leptons, experience a new $SU(2)$ or $U(1)$ gauge
force. Anomaly cancellation enforces the introduction of spectator
quarks so that the top and bottom masses are naturally generated
via a seesaw mechanism. We find the new contributions to the $(S,T,U)$ 
parameters and $\Zbb$ vertex to be generically small. 
We further analyze how the reasonable flavor mixing pattern can be 
generated to ensure the top-seesaw mechanism and
sufficiently suppress the flavor-changing effects for light quarks. 
%An extension to the dynamical symmetry breaking scenario is also given.
Collider signatures for the light Higgs boson and top quark 
are also discussed.
\linebreak
{PACS numbers:\,12.60.-i,\,12.15.-y,\,11.15.Ex~~ 
Phys.Rev.D\,(Rapid Comm, in press)~~ hep-ph/9911266} 
%{[ANL-HEP-PR-99-115  MSUHEP-91015]}
\end{abstract}
%\pacs{PACS number(s): 13.85.Ni  12.60.Fr  14.65.Ha  14.80.Cp}
\vskip1pc]
%]

%\begin{narrowtext}
%\vspace*{-0.8cm}

\setcounter{footnote}{0}
\renewcommand{\thefootnote}{\arabic{footnote}}

%\vspace*{0.25cm}
\noindent
%\underline{\it 1. Introduction} 
%\vspace*{0.2cm}

The single Higgs doublet in the Standard Model (SM) generates
the masses for weak gauge bosons ($W^\pm ,Z^0$) and  
all quarks and leptons by spontaneous 
Electroweak Symmetry Breaking (EWSB). 
However, the striking experimental 
fact is that only the top quark mass 
($m_t=174.3\pm 5.1$\,GeV) lies at the same scale as 
the masses of $(W^\pm ,Z^0)$, 
while all other SM fermions weigh no more than a few GeV. 
This strongly suggests that the top quark sector
may involve certain new gauge dynamics in contrast to all 
light fermions, including the tau lepton.
Following this guideline of model building,
we are forced to introduce new spectator fermions associated with 
the top sector for gauge anomaly cancellation.
We then find that the seesaw mechanism is truely generic to the
top quark mass generation. 

The usual dynamical topcolor scenario \cite{topC} associates
additional strong $SU(3)$ with the top sector, while our topflavor
seesaw models involve either extra $SU(2)$ or $U(1)$ and thus
predict extra color-singlet heavy gauge bosons
such as $W'$ and/or $Z'$. 
The old non-universality\,\cite{old1} or topflavor\,\cite{old2} models
assume the entire third family joins the same extra $SU(2)$ gauge group,
which fails to explain why the top mass is so much larger than 
the tau mass while tau is as light as charm in the second family.
The non-commuting extended technicolor (ETC) model\,\cite{simmons} 
has focused on generating a dynamical $m_t$ by embeding
an extra strong $SU(2)$ into ETC gauge group 
with the anomaly issue ignored for simplicity.  
The recent dynamical topcolor seesaw models\,\cite{Xseesaw,georgi} 
involve an extra singlet heavy quark which is not necessarily
required by the anomaly cancellation since the $SU(3)$
topcolor can be vector-like for SM quarks
and an additional seesaw condition usually needs to be imposed. 
Our construction stresses that a rigorous 
realization of topflavor gauge group of either $SU(2)$ or $U(1)$
in the top-sector (but not tau-sector) 
enforces the introduction of spectator fermions and uniquely leads 
to a seesaw mechanism for $m_t$. 
The topflavor with $SU(2)$ gauge group requires
spectators only in doublet while our topflavor $U(1)$ 
allows either doublet or singlet spectators. The 
doublet spectator fermions always carry weak-isospin so that they
more actively participate in the EWSB dynamics than any singlet
spectator.  The topflavor seesaw scenario with doublet spectator 
fermions thus provides a complementary prospect to the 
original topseesaw idea with extra singlet quark\,\cite{Xseesaw}.
As will be shown below, our new topflavor seesaw models, 
besides theoretically well motivated and defined,
are fully compatible with low energy data
and may further provide exciting collider signatures.
An extension to the dynamical symmetry breaking (DSB)
scenario is also given.

\vspace*{2.5mm}
\noindent
\underline{The Topflavor Seesaw Models}
\linebreak\hspace*{2.5mm}
We construct two types of models in which the top sector, but not
tau sector, experiences a new gauge interaction of $SU(2)_t$ or
$U(1)_t$. The full gauge group is 
${\cal G}_I=SU(3)_c\otimes SU(2)_t\otimes SU(2)_f\otimes U(1)_y$
(called Type-I) or 
${\cal G}_{II}=SU(3)_c\otimes SU(2)_w\otimes U(1)_t\otimes U(1)_f$
(called Type-II). The first two family fermions are singlets under
new $SU(2)_t$ or $U(1)_t$. 
For the third family, a doublet of spectator quarks
$S=(\T,\B)^{T}$ is introduced to 
make the theory free of anomaly (cf. Table\,1). 
A complex Higgs scalar
 $\Phi_I = u+\sigma^0 +i\vec{\tau}\cdot\vec{\chi}$ 
($\Phi_{II} =u+\sigma^0 +i\chi^0$),
with a nonzero vacuum expectation value (VEV) of $u$, 
is introduced to break $\G_I$ ($\G_{II}$) down to the SM gauge group 
$\G_{\rm SM} =$ $SU(3)_c \otimes SU(2)_w \otimes U(1)_y$ at the scale 
$u(\gg\! 246$\,GeV), and then a Higgs doublet 
$H=\(\pi^+,\,(v+h^0+i\pi^0)/\sqrt{2}\)^T$ 
breaks $\G_{\rm SM}$ to the electromagnetic $U(1)_{\rm em}$ at
the scale $v\approx 246$\,GeV.
(Here, $\vec{\tau}$ is the Pauli matrix.)

The gauge sector of Type-I or -II models contain extra massive 
color-singlet weak gauge bosons $(W',Z')$ or $Z'$. The basic parameters
are a small gauge-mixing angle, $\sin\phi$, between heavy and light 
gauge bosons, and a large ratio of two VEVs, $x=u^2/v^2\gg 1$, as
often studied in the literature \cite{old1,simmons,old2,EP}. In fact, 
the $Z'$ of extra $U(1)$ is generic in grand unified models and string
theories \cite{EP}. As long as $\sin\phi$ and $1/x$ are small enough,
all the effects of $W'$ and/or $Z'$  to the
low energy processes can be expressed 
in power expansions of $\sin\phi$ and $1/x$ \cite{old1,simmons,old2}. 
For Type-I models, the true VEV $v_w$ of the EWSB 
is related to the VEV $v$ of
lighter Higgs boson $h^0$ at tree level by
$~
v_w = v\(1-\f{\sin^4\phi}{2x}+O(\f{1}{x^2})\)
$\, ,
with the $W$-boson mass $m_w=g v_w /2$;
while for Type-II models, we have $v_w=v$.
 \\[2mm]
%\begin{table}[H]
%\vspace*{-2.5mm}
%\caption
\noindent
{\small Table.\,1. 
Quantum number assignments for the third family fermions
and the Higgs sector in Type-I and -II models, where
$Q_{3L}=(t_L,b_L)^T$, $L_3=(\nu_{\tau_L}, \tau_L )^T$, 
and $S=(\T,\B)^T$.
}
\vspace*{-2mm}
\begin{center}
\begin{tabular}{c|cccc}
\hline\hline
&&&&\\[-3mm]
~~Type-I~~ & ~$SU(3)_c$~ & ~$SU(2)_t$~ & ~$SU(2)_f$~ & ~$U(1)_y$~~ \\
[0.15cm]\hline\hline
$Q_{3L}$   & \bf{3}        & \bf{2}        & \bf{1}        & $1/3$~~~~\\
$(t_R,b_R)$  & \bf{3}        & \bf{1}        & \bf{1}        & $(4,-2)/3$\\
%$b_R$      & \bf{3}        & \bf{1}        & \bf{1}        & $-2/3$~~~~\\
$S_L$      & \bf{3}        & \bf{1}        & \bf{2}        & $1/3$~~~~\\
$S_R$      & \bf{3}        & \bf{2}        & \bf{1}        & $1/3$~~~~\\
$L_3$      & \bf{1}        & \bf{1}        & \bf{2}        & $-1$~~~~\\
$\tau_R$   & \bf{1}        & \bf{1}        & \bf{1}        & $-2$~~~~\\
\hline
$\Phi$     & \bf{1}        & \bf{2}        & \bf{2}        & $0$~~~~\\
$H$        & \bf{1}        & \bf{1}        & \bf{2}        & $1$~~~~\\
\hline\hline
&&&&\\[-2mm]
~~Type-II~~ & ~$SU(3)_c$~ & ~$SU(2)_w$~ & ~$U(1)_t$~ & ~$U(1)_f$~~ 
\\[0.15cm]
\hline\hline
$Q_{3L}$   & \bf{3}        & \bf{2}        & $1/3$    & $0$~~~~\\
$(t_R,b_R)$  & \bf{3}      & \bf{1}  & $0$   & $(4,-2)/3$~~~~\\
%$b_R$      & \bf{3}        & \bf{1}        & $0$      & $-2/3$~~~~\\
$S_L$      & \bf{3}        & \bf{2}        & $0$      & $1/3$~~~~\\
$S_R$      & \bf{3}        & \bf{2}        & $1/3$    & $0$~~~~\\
$L_3$      & \bf{1}        & \bf{2}        & $0$      & $-1$~~~~\\
$\tau_R$   & \bf{1}        & \bf{1}        & $0$      & $-2$~~~~\\
\hline
$\Phi$     & \bf{1}        & \bf{1}        & $-1/3$   & $1/3$~~~~\\
$H$        & \bf{1}        & \bf{2}        & $0$      & $1$~~~~\\
\hline\hline
\end{tabular}
\end{center}
%\label{Tab:fsu2}
%\end{table}

The main new feature of our models lies in the Yukawa and Higgs sector,
which is the current focus. The scalar $\Phi$ breaks
$SU(2)_t\otimes SU(2)_f$ ($U(1)_t\otimes U(1)_f$) to its diagonal
SM group $SU(2)_w$ ($U(1)_y$) in Type-I(II) models at the scale
$u$($\gg\!\! v$). 
Consequently, it generates the mass of $W'$ and/or $Z'$ as well as
a physical neutral scalar $\sigma^0$. Then,  $\G_{\rm SM}$ breaks
down to $\U1em$ by the doublet Higgs $H$ at the scale 
$v\approx 246$\,GeV and a light neutral Higgs boson $h^0$ is generated.
Therefore, in contrast to the usual two doublet Higgs model (2HDM),
our models have no charged Higgs bosons.  There are a pair of neutral 
scalars $(h^0,\sigma^0)$ with a mixing angle $\alpha$.
The value of $\alpha$ depends on the details of the scalar potential
$V(h^0,\sigma^0)$ and will be treated as a free parameter below.
As a result of the spontaneous symmetry breaking, the scalars are 
expected to obtain tree level masses of the order of their VEV's, 
i.e., $m_h \sim v \sim O(100\,{\rm GeV})$
and $M_\sigma \sim u \sim O({\rm TeV})$.

Defining $\widetilde{H}=-i\tau^2 H^*$, 
from Table~1, we find the following
Yukawa interactions of the third family quarks for both Type-I and -II:
\beq
\ba{l}
       -\dis\f{y_s}{\sqrt{2}}\ov{S_L}\Phi S_R
       -y_{st}\ov{S_L}\widetilde{H}t_R
       -y_{sb}\ov{S_L}{H}b_R
       -\kappa\ov{Q^3_L}S_R \!+\!{\rm h.c.} \\[-5mm]
\ea
\label{eq:LY}  
\eeq
which generate top- and bottom-seesaw mass matrices:
\beq
\ba{l}
-(\ov{t_L},\ov{\T_L})
\left\lgroup\ba{cc} 0      & \kappa \\
                    m_{st} & M_{S}
  \ea\right\rgroup 
\( \ba{c} t_R\\ \T_R \ea \)
                                  \\[3mm]
-(\ov{b_L},\ov{\B_L})\hspace*{-0.2mm}
\left\lgroup\ba{cc} 0      & \kappa \\
                    m_{sb} & M_{S}
  \ea\right\rgroup
\(\ba{c} b_R\\ \B_R \ea \)
+ {\rm h.c.}
\ea
\label{eq:SeesawMass}
\eeq 
where $M_S = y_su/\sqrt{2}$, $m_{st}=y_{st}v/\sqrt{2}$,
$m_{sb}=y_{sb}v/\sqrt{2}$. The parameter
$\kappa$, allowed before spontaneous symmetry breaking,
is expected to be of $O(M_S)$.
Because of the doublet nature of $(\T,\B)$ in our model,
the same $\kappa$ appears in both top- and bottom-seesaw,
in contrast to the recent dynamical seesaw models with singlet
$\chi$ and $\omega$ quarks \cite{Xseesaw,georgi}.
For the parameter space $M_S\gtrsim\kappa \gg m_{st} > m_t$,
the mass eigenvalues of $(t,b)$ and $(\T,\B)$ can be expanded as:
\beq
\ba{l}
m_t =\!\dis \f{m_{st}\kappa}{M_S\sqrt{1+r}}
     \!\[ 1-\f{(m_{st}/M_S)^2}{2(1+r)^2} 
      +O\(\f{m_t^4}{M_S^4}\) \]\!,\\[3mm]
m_b =\!\dis \f{m_{sb}\kappa}{M_S\sqrt{1+r}}
     \!\[ 1-\f{(m_{sb}/M_S)^2}{2(1+r)^2} 
      +O\(\f{m_b^4}{M_S^4}\) \]\!,\\[3mm]
M_{\T}\!=\!\dis M_S\sqrt{1+r}
\!\[\!
1\!+\!\f{z_t^2}{2(1+r)}\!+\!\f{4r+3}{8(1+r)^2} z_t^4\!+\! O(z_t^6)
\!\]\!, \\[3mm]
M_{\B}\!=\!\dis M_S\sqrt{1+r}
\!\[\!
1\!+\!\f{z_b^2}{2(1+r)}\!+\!\f{4r+3}{8(1+r)^2} z_b^4\!+\! O(z_b^6)
\!\]\!, 
\ea
\label{eq:mtbTB}
\eeq    
where  $r\equiv (\kappa /M_S)^2 \sim O(1)$ 
and $z_{t(b)}\equiv m_{t(b)}/\kappa$ with $z_b\ll z_t\ll 1$. 
The mass splitting of the heavy quarks $(\T,\B )$,
 $\Delta M_{\T\B}=M_\T-M_\B $,  is thus deduced as
\beq
\dis\Delta M_{\T\B}
=m_t\[\f{z_t}{2\sqrt{r(1+r)}}+O(z_t^3)\] \ll m_t\,.
\label{eq:MT-MB}
\eeq
The tiny mass-splitting of the $(\T, \B )$ doublet is essential
for satisfying the high precision bound of
$\delta\rho$ or $T$ parameter \cite{peskin}.
    
The seesaw mass matrices in (\ref{eq:SeesawMass}) are diagonalized
by $2\times 2$ bi-unitary transformations,
${K_L^{j\,\dag}}{\cal M}^jK_R^j = {\cal M}^j_{\rm diag}$, 
where the superscript $j\in (t,b)$ specifies the up- and down-type
rotations. The rotation angles $(\theta_L^j,\theta_R^j)$ are
\beq
\ba{l}
\sin\theta_R^j\!=\!\dis 
\f{z_j}{\sqrt{1+r}}\!\[1+\f{r}{1+r}z_j^2\]\!+\!O(z_j^5)\,,\\[3mm]
\sin\theta_L^j\!=\!
\dis\sqrt{\f{r}{1+r}}\!\[1-\f{z_j^2}{1+r}
         -\f{3rz_j^4}{2(1+r)^2}\!+\!O(z_j^6)\].
\ea
\label{eq:angle}
\eeq
Since $z_b/z_t = m_b/m_t \sim 1/40 \ll 1$, the seesaw rotation effects
from the bottom sector are much smaller than that in the top sector. 
If we consider the typical situation with 
$z_b \lesssim O(z_t^2)$, the tiny contribution from the bottom
rotations to $(S,T,U)$ and $R_b$ can be ignored.

With the above seesaw rotations and the $\alpha$-rotation of 
$(h^0,\sigma^0)$ from the Higgs potential, we derive
from (\ref{eq:LY})
the following Yukawa interactions of $(h^0,\sigma^0)$ with $t\bar t$
and $b\bar b$ in the unitary gauge, up to $O(z_t^2, z_b)$,
\beq
\ba{l}
-\dis\f{m_t}{v}\!
\[\!\ca\!\(\!1\!-\!\f{z_t^2}{1+r}\!\)
 \!-\!\sa\!\f{x^{-1/2}}{1+r}\!\(\!1\!-\!\f{1-r}{1+r}z_t^2\!\)\!\]
\!h^0\bar{t}t\\[3mm]
+\dis\f{m_t}{v}\!
\[\!\sa\!\(\!1\!-\!\f{z_t^2}{1+r}\!\)\!
 +\!\ca\!\f{x^{-1/2}}{1+r}\!\(\!1\!-\!\f{1-r}{1+r}z_t^2\!\)\!\]
\!\sigma^0\bar{t}t\\[3mm]
-\dis\f{m_b}{v}\!\[\!\ca\!-\!\sa\!\f{x^{-1/2}}{1+r}\!\]\!
 h^0\bar{b}b
+\dis\f{m_b}{v}\!\[\!\sa\!+\!\ca\!\f{x^{-1/2}}{1+r}\!\]\!
 \sigma^0\bar{b}b
\ea
\label{eq:htt-hbb}
\eeq
where $-\pi/2 \leq \alpha \leq 0$ and 
$(\sa,\,\ca) \equiv (\sin\alpha ,\,\cos\alpha )$. 
Thus, the $ht\bar t$ coupling may be significantly
different from the SM value of $m_t/v_w$ depending on the parameter
space of $(\alpha ,\, x,\, M_S,\, \K)$.  
This may provide, for instance, 
important non-SM signatures via the processes 
$gg\ra h^0 (\ra W^\ast W^\ast \ra \ell\nu\ell\nu )$ 
at the Tevatron, $gg\ra h^0$, $gg \ra \sigma^0 \ra h^0 h^0$
and $WW \ra t\bar{t}$ at the LHC, and
$e^-e^+\ra h^0\ttbar ,\, \nu\bar{\nu} \ttbar$ 
at the high energy linear colliders.

\vspace*{2.5mm}
\noindent
\underline{Constraints from $(S,T,U)$ and $\Zbb$}
\linebreak\hspace*{2.5mm}
The three main new contributions to $(S,T,U)$ and $\Zbb$
arise from (a) the small mixings of heavy $W'$ and/or $Z'$
with $W(Z)$; (b) the $\T$-$t$ and $\B$-$b$ mixings 
from the seesaw mechanism as well as $(\T,\B)$ doublet itself;
(c) the mixing of the Higgs bosons.
The type-(a) contribution is generic to any extended gauge sector
with a breaking pattern $SU(2)_1\otimes SU(2)_2\ra SU(2)_w$
or $U(1)_1\otimes U(1)_2\ra U(1)_y$ and can safely fit the data as long
as the mixing angle $\sin\phi$ and the ratio $1/x$ are small 
enough \cite{old2,EP}. Our real concern 
is the new type-(b) and -(c) corrections.
The usual expectation is that only
$SU(2)_w$ {\it singlet} heavy fermions are phenomenologically safe
\cite{Xseesaw}, but our analysis shows that the contributions of the
doublet fermions $(\T,\B)$ in our seesaw mechanism 
are also generically small enough to agree with the current data.
For simplicity, we compute the type-(b) contributions up to $O(z_t^4)$
in seesaw expansion while keeping leading orders in small 
$\sin\phi$ and $1/x$ expansions \cite{old2,EP}. 
To the leading order in $\sin\phi$ and $1/x$, the doublet  
$(\T,\B)$ behaves essentially {\it vector-like} under the SM 
gauge group, and thus their heavy masses are expected to respect the 
decoupling theorem \cite{vectorF}.
Even though the masses $M_S$ and $\K$ are invariant under $\GSM$,
the other seesaw mass terms ($m_{st}$ and $m_{sb}$) are not.
It is a nontrivial task to confirm that the spectator-fermion
corrections to $(S,T,U)$ can decouple sufficiently since
the fermion-loops involving heavy $\T / \B$ do contribute 
dangerous $O(M_S^2)$ and $O(M_S^0)$ terms to the 
self-energies of $W/Z$.

The calculations of $(S,T,U)$ are tedious,
but the results to $O(z_t^2,z_b)$ can be compactly summarized,
\beq
\ba{l}
S\!=\!\dis\f{4N_c}{9\pi}\!\[\ln\!\f{M_\T}{m_t}\!-\!\f{7}{8}
\!+\!\f{1}{16h_t}\!-\!\f{1}{560h_t^2}\]\!\f{z_t^2}{1+r} \,,\\[3mm]
T\! =\!\dis\f{N_c h_t}{16\pi s_w^2c_w^2}\!\[8\ln\!\f{M_\B}{m_t}\!
  +\!\f{4}{3r}\! -\! 6\! \;\]\!\f{z_t^2}{1+r} \,,
\\[3mm]
U\!=\!\dis\f{N_c}{6\pi}\[1+\f{1}{10h_t}
      +\f{1}{70h_t^2}\]\f{z_t^2}{1+r} \,,
\ea
\label{eq:STU}
\eeq
where $h_t\equiv  (m_t/m_z)^2$ and
we have ignored tiny $O(1/h_t^3)$ terms inside $[\cdots ]$.  
%$T_0=N_c h_t/(16\pi s_w^2c_w^2)$ 
%is the SM top quark contribution to $T$. 
($s_w\equiv \sin\theta_W$ and $\theta_W$ is the weak mixing angle.)
We see that these new contributions are 
phenomenologically safe since $z_t^2\ll 1$. For instance, taking
$M_S=2\K=5$\,TeV, we have 
$\STU =(4\times 10^{-3}$, $0.13,\,6\times 10^{-4})$,
while choosing $(M_S,\K)=(5,\,4)$\,TeV, we get
$\STU =(1.4\times 10^{-3},\,0.04,\,2\times 10^{-4})$.
So, the seesaw corrections to $(S,U)$ are generally negligible.

There are also contributions to $\STU$ from the Higgs bosons.
In the limit of $1/x \ll 1$, the heavy $\sigma^0$ only indirectly
couples to $W/Z$ via its $\alpha$-mixing with the light $h^0$.
The interactions of $(h^0, \sigma^0)$ with $(W^\pm ,Z^0)$ are
\beq
\ba{l}
\dis
\f{e^2}{2 s_w^2}    W_{\mu}^2 
(c_\alpha h^0\!+\!s_\alpha\sigma^0)^2+
\f{e^2}{2 s_w^2 c_w^2}Z_\mu^2 
(c_\alpha h^0\!+\!s_\alpha\sigma^0)^2 +
\\[3.3mm]
\dis
\f{e m_w}{s_w}  W_{\mu}^2 
(c_\alpha h^0\!+\!s_\alpha\sigma^0) +
\f{e m_z}{s_w c_w}Z_\mu^2 
(c_\alpha h^0\!+\!s_\alpha\sigma^0)\,.
\ea
\eeq
Thus, the $(h^0, \sigma^0)$ contribute to low energy observables
in the same way as the SM Higgs, but with a scaling factor of 
$(c^2_\alpha ,s^2_\alpha )$.
From the SM Higgs correction to $\STU$\,\cite{peskin}, 
we derive the additional contributions from $(h^0, \sigma^0)$:
\beq
\ba{l}
  \Delta S\!=\!\dis\f{1}{12\pi}\!
      \[ c^2_\alpha \ln\!\f{m^2_h}{m_z^2}\!
         - \ln\!\f{(m^2_h)^{\rm sm}_{\rm ref}}{m_z^2}\!
      + s^2_\alpha \ln\!\f{M^2_\sigma}{m^2_z}\!\]
    \, , \\[3mm]
 \Delta T\! =\!\dis\f{-3}{16\pi c_w^2}\!
     \[ c^2_\alpha \ln\!\f{m^2_h}{m_z^2}\!
        - \ln\!\f{(m^2_h)^{\rm sm}_{\rm ref}}{m_z^2}\!
     + s^2_\alpha \ln\!\f{M^2_\sigma}{m^2_z}\!\]\,,
% \\[3mm]
%U\!\simeq\!0\, ,
\ea
\label{eq:HiggsSTU}
\eeq
and $\Delta U\!\simeq\!0$, where
$(m_h)^{\rm sm}_{\rm ref}$ is the reference value of the SM Higgs mass.
For $(m_h)^{\rm sm}_{\rm ref}=m_h=100$ GeV and 
$M_\sigma=1$\,TeV,  we find 
$(\Delta S,\Delta T)=(0.02,-0.07)$ with $s^2_\alpha=0.2$.  
The contributions of the Higgs and seesaw sectors
to $T$ can have opposite signs, which makes our model easily
accommodate the data with a small $T$ for reasonable $(M_\sigma ,M_S)$.
On the other hand, $M_S$ is bounded from 
above (since a larger $M_S$ lifts up $S$ to positive side)
and also from below (since a light $M_S$ pushes both $S$ and $T$ 
towards negative).   
Considering $1/x\ll 1$ and summing up dominant contributions
in the Higgs and seesaw sectors, we can derive constraints in the 
$(M_\sigma ,M_S)$ plane from the precision fit of $(S,T)$\,\cite{EPSTU}, 
as shown in Fig.\,1. We have chosen 
$(m_h)^{\rm sm}_{\rm ref}=100$\,GeV for the $(S,T)$ fit, with
the complete 1-loop SM corrections included (in accord
with the precision of our 1-loop new physics results). 
The fitted values of   
$(S,T) = (0.13\pm 0.11, -0.13\pm 0.14)$
deviate from $(0,0)$ at $1\sigma$ level.
Fig.\,1 shows that $M_\sigma$ is always 
bounded from below since a too light $\sigma^0$
drives both $(\Delta S,\Delta T)$ to zero.\\[-12mm] 
%\begin{figure}
\hspace*{-0.4cm}
\epsfig{file=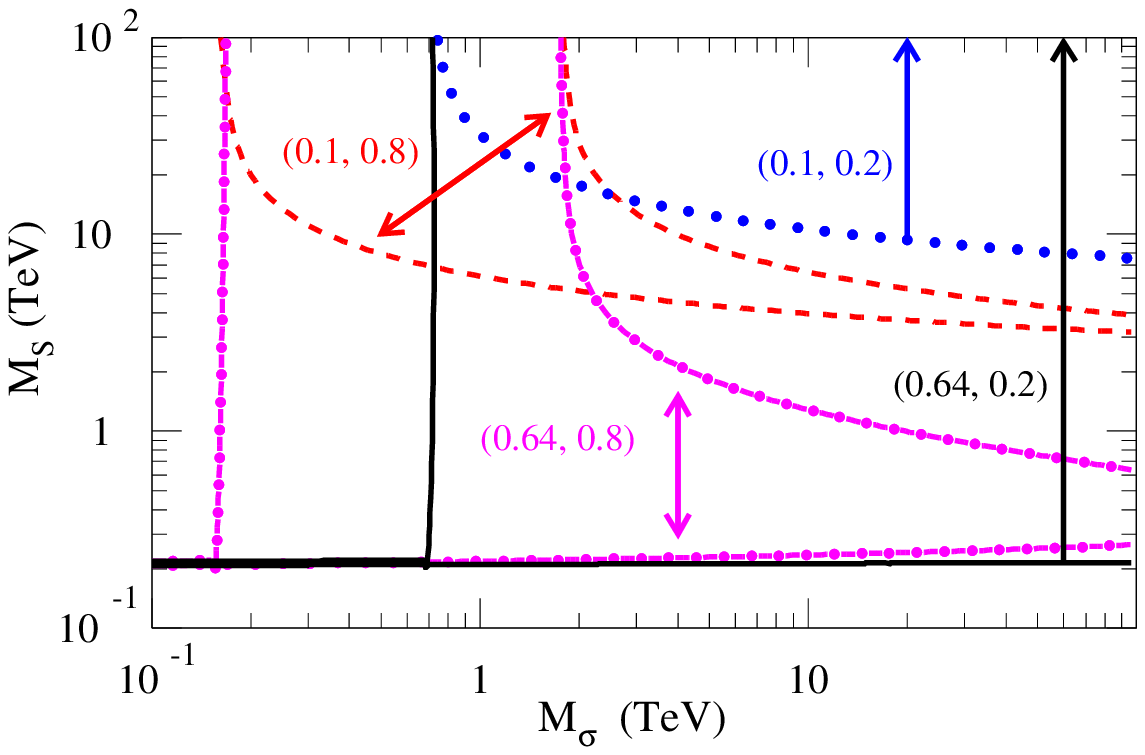,width=10cm,height=7.5cm}\\[-7mm]
%\centerline{\epsfbox{msms.eps}}
%\caption{
\hspace*{3mm}
Fig.\,1. {\small Constraints on $(M_\sigma ,M_S)$  
by ($S$,\,$T$) fit at $95\%$\,C.L., 
for $m_h=100$\,GeV and four sets of ($r,\,s^2_\alpha$) as shown.  
The allowed regions (indicated by arrows) lie between 
two lines (or above one line) appropriate to each parameter set.}
%}
%\label{fig:STexclude}
%\end{figure}

We finally discuss the ratio of $Z$-decay width
$R_b=\Gamma (Z\!\ra\!b\bar{b})/\Gamma (Z\!\ra\!{\rm hadrons})$ 
and the $\Zbb$ coupling asymmetry 
$A_b=(g_{bL}^2\!-\!g_{bR}^2)/(g_{bL}^2\!+\!g_{bR}^2)$. The current 
experimental data from $R_b$ and $A_b$ can be translated
into the bounds on the allowed deviation of the $\Zbb$-couplings
$(g_{bL},g_{bR})$ from their SM values,
$0.002\leq \delta g_{bL} \leq 0.009$
and $0.004\leq \delta g_{bR} \leq 0.036$, 
at $2\sigma$ level \cite{RbAb}. 
It is straightforward to compute the corrections to 
$\Zbb$ couplings from the seesaw sector of our
model. The correction associated with the
top sector only comes from loop and is of
$O(z_t^2)$ %in the leading order 
so that it is generally small,
but the bottom-seesaw induces a tree level
correction $\delta g_{bR}^{\rm new}$ to the right-handed 
$\Zbb$ coupling,
\beq
\dis \delta g_{bR}^{\rm new} =-\f{e}{2s_wc_w}
(\sin\theta_R^b)^2~.
\eeq
This negative correction is at the order of
$(\sin\theta_R^b)^2\simeq (m_b/\K)^2/(1+r) \lesssim 
O(10^{-6}-10^{-8})$ for $\K=O(1-10)$\,TeV
and thus essentially negligible. 
This feature is different from the recent dynamical 
seesaw models with singlets $\chi$ and $\omega$ in which
the left-handed (instead of right-handed) 
$b$-$\omega$ mixings contribute to $\Zbb$ vertex \cite{georgi}.
Another nice feature is that our models contain
no charged Higgs and are thus free of their 
undesirable negative correction to $R_b$
and also their enhancement to $b \rightarrow s \gamma$
decay rate in the usual 2HDM \cite{RbAb}.

\vspace*{2.5mm}
\noindent
\underline{Quark Mass Matrices and Flavor Mixings}
\linebreak\hspace*{2.5mm}
To establish realistic flavor mixings among all three
families with the well constrained 
Cabibbo-Kobayashi-Maskawa (CKM) matrix generated
is a more challenging task. 
We do not want to spoil the seesaw pattern of
the mass matrices in (\ref{eq:SeesawMass}) after the
mixings with the first two family fermions are included,
and we also need to properly suppress flavor-changing effects 
associated with the light quarks.
The quantum number assignments in Table\,1 do not automatically
suppress the mixings of $(\T,\B)$ and $(t,b)$ with light fermions. 
We impose a simple discrete $Z_4$ symmetry
to ensure the desired pattern of the $4\times 4$ mass matrices for
up- and down-type quarks. 
%%% We will not discuss the origin of such a discrete $Z_4$ for 
%%% current effective theory study at weak scale, but assume it to 
%%% be a residue of flavor symmetry breaking at a much higher scale.
%%This $Z_4$ may be a natural consequence 
%%of breaking the flavor symmetry at a higher energy scale.
Under $Z_4=\exp (in\pi/2)$ with
$n\in (0,1,2,3)$, we define the following field transformations:
\beq
\ba{c}
Q_{3L}\ra iQ_{3L},~~ S_R\ra iS_R, ~~\Phi\ra i\Phi, \\
(t_R,b_R)\ra -(t_R,b_R),~~ S_L\ra -S_L,
\ea
\label{eq:Z4}
\eeq
and other fields are unchanged by $Z_4$.
Then, we can write down all relevant effective operators in the
quark Yukawa sector, invariant under $\G_I(\G_{II})$.
For instance, in the Type-I models,
the Yukawa Lagrangian $-{\cal L}_{Y(U)}$ 
of the up-type quarks becomes
\beq
\ba{l}
\dis\hspace*{-2mm}
\sum_{i,j=1}^2\!\! 
y_{ij}\ov{Q_{iL}}\HH  u_{jR} \!+
y_{i3}\ov{Q_{iL}}\HH  t_R\f{\det\Phi}{\cut_f^2} \!+ 
\f{y_{i4}}{\sqrt{2}}\ov{Q_{iL}}\Phi S_R\f{\det\Phi}{\cut_f^2}
\nonumber \\
\dis
\,+y_{3j}\ov{Q_{3L}}\Phi^\dag\HH u_{jR}\f{\det\Phi}{\cut_f^3}\!+
y_{33}\ov{Q_{3L}}\f{\Phi^\dag}{\cut_f}\HH t_R 
\!+\!\K\ov{Q_{3L}}S_R \nonumber  \\
\dis
\,+y_{4j}\ov{S_L}\HH u_{jR}\f{\det\Phi}{\cut_f^2} +
y_{ts}\ov{S_L}\HH t_R +\f{y_S}{\sqrt{2}}\ov{S_L}\Phi S_R\\[-8mm]
\ea
\label{eq:LY_U}
\eeq
where $\Lambda_f$ is the cutoff scale of the flavor symmetry breaking. 
Defining $m_{ij} =y_{ij}v/\sqrt{2}$, 
we find that the resulting mass matrix for $(u,c,t,\T)$  
poses a natural hierarchy, 
\beq
{\cal M}_{u}~=~ 
\(\ba{cccc}
m_{11} & m_{12}   & m_{13}\epsilon^2 & m_{14}\delta \\[1.5mm]
m_{21} & m_{22}   & m_{23}\epsilon^2 & m_{24}\delta \\[1.5mm]
m_{31}\epsilon^3  & m_{33}\epsilon^3 & m_{33}\epsilon 
                     & \kappa \\[1.5mm]
m_{41}\epsilon^2  & m_{42}\epsilon^2  & m_{st} & M_{S}   
 \ea 
\)
\label{eq:MU44}
\eeq
in which $\epsilon =u/\Lambda_f$ and $\delta =\epsilon^2(u/v)$
are small parameters. The down-type quarks exhibit
a similar pattern in ${\cal M}_d$.
A proper bi-unitary field transformation, containing the dominating
$2\times 2$ seesaw-rotations in the $t$-$\T$ ($b$-$\B$) sector,
can first rotate away the small mixings of $\T (\B)$ with all light 
quarks so that the $4\times 4$ mass matrix reduces to $3\times 3$
for the three-family quarks of the SM, i.e.,
\beq
\widehat{\cal M}_{u}~=~ 
\(\ba{ccc}
  m_{11} & m_{12} & m_{13}'\delta  \\[1.5mm]
  m_{21} & m_{22} & m_{23}'\delta  \\[1.5mm]
  m_{31}'\epsilon^2  & m_{32}'\epsilon^2  & m_t' 
 \ea 
\)
\label{eq:MU33}
\eeq
where $m_t'\lesssim m_t$. 
A similar analysis applies to Type-II models.
Following the procedure of Ref.\,\cite{HY-prl}, realistic CKM
mixings of SM fermions can be generated with a proper construction
of left-handed rotations for the up- and down-type quarks. 
The flavor changing effects associated with light quarks were
found to be reasonably suppressed\,\cite{HY-prl} 
in consistency with low energy data, while
right-handed mixings are constrained 
by the mass pattern (\ref{eq:MU33}).
Sizable flavor mixings between right-handed
$c_R$ and $t_R$ are allowed \cite{HY-prl}: 
\beq
\dis K_{UR}^{tc}\lesssim \sqrt{1-(m_t'/m_t)^2} \simeq 0.11-0.33\,,
\label{eq:tc-mixing}
\eeq
for reasonable values of $\delta m_t=m_t-m_t'=O(1-10)$\,GeV.
Hence, the charm-gluon fusion process 
$gc\ra h^0t$ \cite{yuan} provides an important Higgs discovery 
channel at the LHC.

\vspace*{2.5mm}
\noindent
\underline{Extension to Dynamical Symmetry Breaking Scenario }
\linebreak\hspace*{2.5mm}
While the above topflavor seesaw models have provided the crucial
ingredients on how a large top mass is generated together with the
EWSB, it is desirable to invoke dynamical symmetry breaking 
at the TeV scale without introducing fundamental Higgs.
Here, we only consider the {\it simplest} DSB realization
of our seesaw mechanism of Type-II models, which is called 
Type-IID below. 

To replace the fundamental VEV $\langle H\rangle$ by a dynamical
condensate, we may introduce a
strong $SU(3)_t$ gauge interaction for $(t_R,\,b_R)$ and ${\cal S}_L$
while all other quarks join the weaker $SU(3)_f$ group.
(The strong $SU(3)_t$  is traditionally called topcolor\,\cite{topC}.)
Thus, our Type-IID models, 
as an extension of the above Type-II scenario, 
have the gauge structure
$\G_{IID}=SU(3)_t\otimes SU(3)_f\otimes SU(2)_w\otimes 
          U(1)_t\otimes U(1)_f$,  
which turns out to match the gauge group of the original 
non-seesaw topcolor models\,\cite{topC}.
But our Type-IID models differ in that they contain new doublet
spectator fermions for generating the seesaw mechanism and have
very different quantum number arrangement 
enforced by the anomaly cancellation (cf. Table.\,2).
The first two family fermions are charged under weaker
$SU(3)_f$ and $U(1)_f$ as in the SM. The strong $U(1)_t$
is now designed to tilt the vacuum such that only top but not bottom 
gets a large seesaw mass, cf. (\ref{eq:NJL})-(\ref{eq:tilt}). 

The gauge group $\G_{IID}$ first breaks down to   
$~\GSM$ at the scale $u$ and 
then breaks down to $U(1)_{\rm em}$ at the scale $v$.
The first step breaking may be effectively parametrized by a scalar
$\Phi$ with VEV $u$, from which the massive octet colorons
($G_\mu^{\prime a}$) and $U(1)$ gauge boson ($Z'_\mu$) are 
generated at the scale $M_c\sim M_y\lesssim 4\pi u$. ($M_c$ and
$M_y$ are the masses of $G_\mu^{\prime a}$ and $Z'_\mu$, respectively.)
Thus, integrating out the heavy $G_\mu^{\prime a}$ and $Z'_\mu$
results in the effective interaction for the third family quarks:
\beq
\ba{l}
\dis
-\f{4\pi\K_c}{M_c^2}\(
\ov{S_L}\gamma^\mu \GL S_L +
\ov{t_R}\gamma^\mu \GL t_R +
\ov{b_R}\gamma^\mu \GL b_R \)^2\\
\dis
-\f{4\pi\K_y}{M_y^2}\(
\f{1}{6}\ov{S_L}\gamma^\mu S_L +
\f{2}{3}\ov{t_R}\gamma^\mu t_R -
\f{1}{3}\ov{b_R}\gamma^\mu b_R \)^2 .
\ea
\label{eq:Leff-VV}
\eeq
Here, $(\K_c,\K_y)=(g_3^2\cot^2\theta ,\,g'^2\cot^2\theta')/8\pi$,
with $g_3$($g'$) the gauge coupling of the SM color (hypercharge)
force and $\theta$($\theta'$) the mixing angle of 
the two $SU(3)$'s ($U(1)$'s)\,\cite{topC}.
Applying Fierz transformation to (\ref{eq:Leff-VV}) leads to
Nambu-Jona-Lasinio\,(NJL) type interactions, 
for $M_y\simeq M_c$,
\beq
\ba{l}
\dis
\hspace*{-2mm}
\f{8\pi}{M_c^2}\!\!\[\!
(\bar{\K}_c\!+\!\f{2\K_y}{9N_c})(\ov{S_L}t_R)(\ov{t_R}S_L)\!+\!
(\bar{\K}_c\!-\!\f{\K_y}{9N_c})(\ov{S_L}b_R)(\ov{b_R}S_L)
\!\] \\[-2mm]
\ea
\label{eq:NJL}
\eeq
where $\bar{\K}_c=\K_c(1-1/N_c^2)$. 
In the large-$N_c$ expansion, a generic NJL-type vertex,
${\widehat{\K}}{\Lambda^{-2}}\!\(\ov{X_L}Y_R\)\!\(\ov{Y_R}X_L\)$,
has a critical coupling 
$\widehat{\K}_{\rm crit}\simeq 8\pi^2/N_c$ 
for the dynamical condensation.
With the $U(1)_t$-tilting in (\ref{eq:NJL}), we thus have 
$\langle\ov{\T_L}t_R \rangle\neq 0$ and
$\langle\ov{\B_L}b_R \rangle = 0$, provided 
\beq
\dis
\f{3\pi}{8}-\f{\K_y}{12}\leq \K_c \leq \f{3\pi}{8}+\f{\K_y}{24}~.
\label{eq:tilt}
\eeq
An essential feature of our scenario is that 
the spectator $\T_L$, but not the SM $t_L$, plays 
the key role in the dynamical condensate which generates the EWSB and
seesaw top-mass, in contrast to the recent topseesaw models 
involving extra singlet heavy quark\,\cite{Xseesaw,georgi}.
Consequently, two composite Higgs doublets $H_{st}$ and $H_{sb}$
are generated, which are made of $(\ov{S_L}t_R)$ and $(\ov{S_L}b_R)$,
respectively.  The $U(1)$-tilting in (\ref{eq:tilt}) ensures
that $\langle H_{st}\rangle\neq 0$ and $\langle H_{sb}\rangle =0$.
Thus, the Higgs spectrum contains a top-condensate
Higgs $h_{st}^0$, a $b$-Higgs $h_{sb}^0$ and three $b$-pions
$(\pi_{sb}^0,\pi_{sb}^\pm )$, as hybirds between $(\T_L,\B_L)$ and
$(t_R,b_R)$.
With the coloron mass $M_c\lesssim 4\pi u$ as a cut-off, 
we can now re-derive Pagels-Stokar formula
for generating both the dynamical top mass and EWSB
with $v\approx 246$\,GeV, i.e.,
\beq
v^2 = \dis\f{N_c}{8\pi^2}\f{m_t^2}{\sin^2\theta^t_L} 
\ln\f{M_c^2}{M_S^2(1+r)}+O(z_t^2)\,,
\label{eq:PS}
\eeq 
where, for example, $(\K,M_S,M_c )\sim (2,5,50)$\,TeV 
and $m_t/\sin\theta^t_L\sim 600$\,GeV. Note that (\ref{eq:PS})
involves the left-handed (instead of right-handed) 
seesaw rotation angle $\theta_L^t$,
unlike the situation in Refs.\,\cite{Xseesaw,georgi}. 

%\begin{table}[ht]
%\vspace*{-5mm}
%\caption{\small 
{\small Table\,2.
Quantum number assignments for the third family fermions and 
the effective Higgs scalar $\Phi$ in Type-IID models.
}
\vspace*{-4mm}
\begin{center}
\begin{tabular}{c|ccccc }
\hline\hline  % ---------------------------------------------------------
&&&&&\\[-0.2cm]
Type-IID &  
{ $SU(3)_t$} & 
{ $SU(3)_f$} & 
{ $SU(2)_w$} & 
{ $U(1)_t$} & 
{ $U(1)_f$} \\
[0.15cm]\hline\hline
&&&&&\\[-0.25cm]
$Q_{3L}$ & \bf{1} & \bf{3} & \bf{2} & $0$ & $1/3$
\\ %[1.3mm]
$(t_R,b_R)$ & \bf{3} & \bf{1} & \bf{1} & $(4,-2)/3$ & $0$
\\ %[1.3mm]
$S_L$ & \bf{3} & \bf{1} & \bf{2} & $1/3$ & $0$
\\ %[1.3mm]
$S_R$ & \bf{1} & \bf{3} & \bf{2} & $0$ & $1/3$
\\%[1.3mm]
$L_3$ & \bf{1} & \bf{1} & \bf{2} & $-1$ & $0$
\\%[1.3mm]
$\tau_R$ & \bf{1} & \bf{1} & \bf{1} & $-2$ & $0$ \\[1.3mm]
\hline %\hline
&&&&\\[-0.35cm]
$\Phi$ & \bf{3} & $\overline{\bf 3}$ & \bf{1} & $1/3$ & $-1/3$
\\ [1mm]
%\\[1mm]  %[0.2cm]
\hline\hline % ---------------------------------------------------
\end{tabular}
\end{center}
%\label{Tab:QNo-DSB2}
%\end{table}

As a final remark, 
the small masses of $b$, $\tau$ and the first two family
fermions have to be generated by different mechanisms, which are
much more model-dependent\,\cite{tseesaw2}.  For instance,
they can come from higher dimensional effective operators \cite{georgi},
composite Higgs doublet (formed at higher scale) 
with a small VEV $v_f\!\gtrsim \!O(1\!-\!10)$\,{\rm GeV} \cite{DS},
or extended technicolor interactions\cite{ETC}.

%A different way of solving the naturalness problem is to supersymmetrize
%our topflavor seesaw models (especially Type-II and its variations) 
%with fundamental Higgs bosons. Work along this line is in progress.

\smallskip 
We thank B. A. Dobrescu, J. Erler, A. Grant, C.T. Hill, 
and M. E. Peskin for valuable discussions, 
and G. Cvetic and E. H. Simmons for helpful comments on the manuscript.
This work is supported by the U.S. NSF (PHY-9802564) and 
DOE (Contract W-31-109-Eng-38).
\\[-9mm]

%\end{narrowtext}
\end{document}